\documentclass[compress,twocolumn,5p,sort]{elsarticle}
\usepackage[utf8]{inputenc}
\usepackage{amsmath}
\usepackage{amsfonts}
\usepackage{amssymb}
\usepackage{graphicx}
\usepackage{float}
\usepackage{stfloats}
\usepackage{color}
\usepackage{fancyhdr}
\usepackage{booktabs}
\usepackage{multirow}
\usepackage{threeparttable}
\usepackage{pdflscape}
\usepackage{mathrsfs}
\usepackage{ulem}
\usepackage{textcomp}
\usepackage{epstopdf}
\usepackage{gensymb}
\usepackage[switch]{lineno}

\graphicspath{{figure/}}

\journal{ }

\begin{document}
	\begin{frontmatter}
		
		\title{Revealing mechanism of pore defect formation in laser directed energy deposition of aluminum alloy via in-situ synchrotron X-ray imaging}
		
		\author[AECC]{Wei. Liu\corref{contrib}}
		
		\author[IHEP]{Yuxiao Li\corref{contrib}}
		
		\author[IHEP]{Chunxia Yao}
		
		\author[IHEP,UCAS]{Dongsheng Zhang}
		
		\author[IHEP]{Darui Sun}
		
		\author[CAEP]{Sen Chen}
		
		\author[AECC]{Yu Wu}
		
		\author[SSRF]{Jun Wang}
		
		\author[SWJTU]{Lei Lu}
		
		\author[SWJTU]{Sheng-Nian Luo\corref{corr-author}}
		\ead{sluo@swjtu.edu.cn}
		
		\author[IHEP]{Ye Tao\corref{corr-author}}
		\ead{taoy@ihep.ac.cn}
		
		\author[IHEP,UCAS]{Bingbing Zhang\corref{corr-author}}
		\ead{zhangbb@ihep.ac.cn}

		\address[IHEP]{Institute of High Energy Physics, Chinese Academy of Sciences, Beijing, China}
		
		\address[UCAS]{University of Chinese Academy of Sciences, Beijing, People's Republic of China}
		\address[AECC]{3D Printing Research and Engineering Technology Center, Beijing Institute of Aeronautical Materials, Beijing, China}

		\address[SWJTU]{School of Materials Science and Engineering, Southwest Jiaotong University, Chengdu, Sichuan, People's Republic of China}
		
		\address[CAEP]{Institute of Fluid Physics, China Academy of Engineering Physics, Mianyang, Sichuan, China}
		
		\address[SSRF]{Shanghai Synchrotron Radiation Facility, Shanghai Advanced Research Institute, Chinese Academy of Sciences, Shanghai, China}

		\cortext[corr-author]{Corresponding author}
		\cortext[contrib]{Authors contributed equally}

		\begin{abstract}
			Laser metal additive manufacturing technology is capable of producing components with complex geometries and compositions that cannot be realized by conventional manufacturing methods. However, a large number of pores generated during the additive manufacturing process greatly affect the mechanical properties of the additively manufactured parts, and the mechanism of such pore generation has not been revealed by direct observation clearly. Here, we report the mechanism of pore generation in the laser direct energy deposition process as revealed by {\it in-situ} high-speed high-resolution synchrotron X-ray imaging. We found that dissolution and re-precipitation of external gases and precipitation of metal vapors are the two main mechanisms of pore formation. We further explored the effects of different process parameters on the generation of pores and optimized the process to suppress pore generation. This work provides important insights into the formation of porosity defects during laser metal additive manufacturing, and can provide guidance for related process optimization.
			
		\end{abstract}
		
		
	\end{frontmatter}
	
	\section{Introduction}
	
	Additively manufactured (AM) aluminum alloy parts have a promising application in lightweight structures in the automotive, marine, and aerospace fields\cite{martin20173d,croteau2018microstructure}. One of the main methods of additive manufacturing is laser direct energy deposition (DED)\cite{croteau2018microstructure,debroy2018additive}, which can be used to fabricate complex geometries as well as to repair such as aerospace blade\cite{piscopo2022current}. In the DED process, a high-power ($\sim$300--2000 W) and highly focused (spot size $\sim$50--500 \micro m) laser is scanned at a certain speed\cite{smith2016linking}, and the powder is simultaneously transported to the focused position of the laser beam on the substrate, resulting in the formation of a molten pool. Then a fully dense part is deposited layer by layer in a continuous process. However, the use of DED in the aforementioned fields is constrained by the porosity introduced during the manufacturing process, which can adversely affect the final mechanical properties of the component\cite{sterling2016fatigue,kim2022characterization}.

	A melt pool is formed at the top of the substrate during the DED process, and a large number of bubbles precipitate inside due to a variety of factors. As the scanning progresses, these bubbles are continuously captured by the solidification edges and remain inside the material, forming a porosity defect\cite{zhang2024pore}. Such defects have been observed in a variety of DED forming materials, such as titanium alloys and aluminum alloys. According to previous research, these pores can be broadly classified into three categories according to the cause of their formation. First, the hydrogen pores (e.g., hydrogen from the decomposition of non-dense substrates and residual moisture)\cite{svetlizky2021directed,xu2024towards}. And second, the laser vaporizes the metal, then the recoil pressure pushes the molten metal away from the laser-material interaction zone. As the laser power increases, the recoil pressure is large enough to open a deep vapor depression called a keyhole, which is unstable and prone to collapse due to various perturbations within the molten pool, thus separating metal vapor bubbles\cite{chen2023situ,wolff2021situ}. The third category is the cavity due to the difference in the volume of the solid and liquid phases during solidification and shrinkage\cite{liu2021review}. These pores remaining in the final part may become stress concentration points and crack initiation, which may cause damage to fatigue life and other mechanical properties\cite{tammas2017influence,jost2021effects,martin20173d}. Conventional porosity studies are usually performed with {\it ex-situ} non-real-time diagnosis techniques such as electron microscope scanning, etc\cite{pang2019characterisation,kistler2019effect}, which can not capture the phenomena of pore formation and growth. To comprehensively study the mechanism of pore formation and evolution, it is essential to employ {\it in-situ} real-time observation techniques.

	Synchrotron X-ray technology enables {\it in-situ} real-time observation of the additive manufacturing process. Researchers have captured some dynamic features of the melt pool pores, including the keyhole evolution\cite{cunningham2019keyhole}; the formation of pores during laser powder bed fusion\cite{martin2019dynamics}; the elimination of pores\cite{hojjatzadeh2019pore}; the migration  of pores under Marangoni flow or the formation of hydrogen pores\cite{leung2018situ}; and the evolution of pores during the multi-layer molding process\cite{sinclair2020situ}, etc. However, for the DED technique, the larger melt pool and the protective gas or powder shock perturbations lead to a complex pore evolution. Zhang et al. used synchrotron X-ray imaging to investigate the pore migration and pore coalescence of DED\cite{zhang2024pore}. However, these studies lack further exploration of the mechanism of bubble formation and evolution, the synergistic effect of different mechanisms on bubble evolution remains to be understood, and the optimization strategy of DED process under real production conditions needs further research assistance.

	In order to better understand the formation mechanism of internal defects, numerical simulation methods can supplement the experiment and analyze more influencing factors. Some simulation models have initially revealed the pore evolution during the additive process. Some model numerically investigated pore dynamics. There are also models that investigate the physical processes of keyhole formation, revealing the interactive effects of recoil pressure, surface tension, and Marangoni convection, as well as the competing effects of gravity, drag, buoyancy, and thermocapillary forces on bubble motion in melt pool\cite{khairallah2016laser,tan2014analysis,khairallah2020controlling,wang2020evaporation}. However, the existing multiphysics coupled computational models mainly study the physical characteristics (e.g., melt pool flow or temperature field). As for the bubble evolution process, a large number of studies have been carried out for the keyhole mechanism\cite{shrestha2019numerical,wang2022mechanism}, and fewer simulations have been carried out for other types of mechanisms\cite{zhang2024pore}, such as hydrogen holes. At the same time, such methods cannot analyze the microscopic process. Therefore, it is necessary to conduct a more comprehensive study on the pore dynamics during the AM process. It is necessary to reveal the characteristics of pore and melt pool dynamics in DED by combining multiscale simulation and {\it in-situ} X-ray imaging experiments.

	\begin{figure*}[htbp]
		\center
		\includegraphics[width=\linewidth]{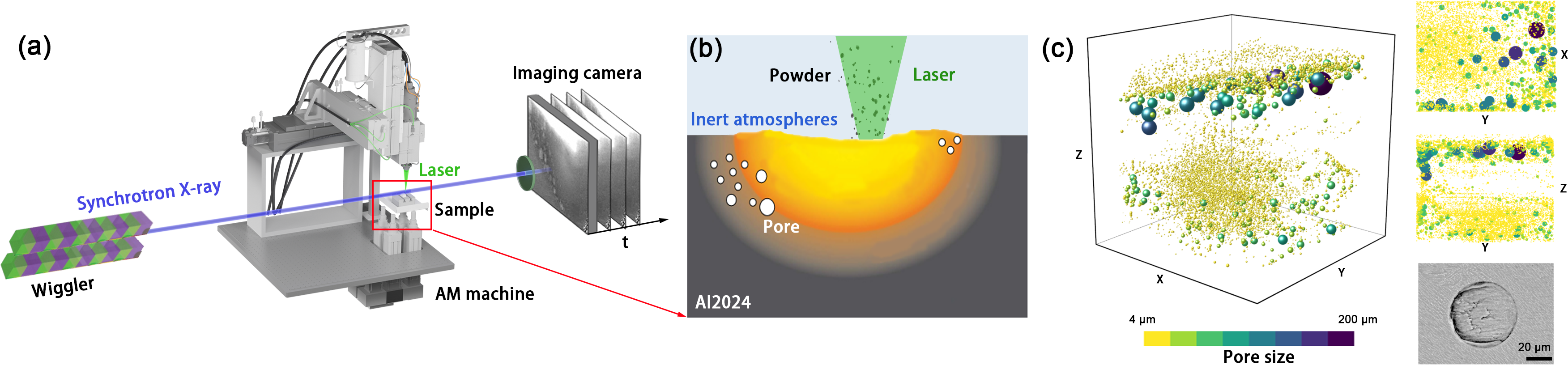}
		\caption{(a) Schematic of laser directed energy deposition with time-resolved X-ray imaging. (b) Illustration of melting pool. (c) CT characterization of an LDED-fabricated sample.}
		\label{fig_1}
	\end{figure*}

	Here, we performed {\it in-situ} synchrotron X-ray imaging of the DED process in the commercial aluminum alloy Al2024 (courtesy of Beijing Institute of Aeronautical Materials), which has important applications in the aerospace and automotive industries, in order to investigate the mechanisms of pore evolution during the DED process. We quantified the pore evolution behavior in the imaging data and identified several major types of pore formation mechanisms, including dissolution and precipitation of ambient gases, formation of metal vapors, and volume collapse during solidification shrinkage. We observed the interaction of dissolved gases and metal vapors on the volume deformation and kinematic evolution of pores. And the porosity under different laser strategies was investigated based on the {\it in-situ} experimental results. Meanwhile, we utilized a combination of multiphysics coupled computational models and molecular dynamics (MD) simulation to assist in verifying the above possible mechanisms, and explored the effects of different laser parameters on pores formation and evolution. This work will help us to further understand the pores formation and evolution mechanism, and assist in the optimization of DED process strategy.

	\section{Results}
	\noindent \textbf{Experimental setup and ex-situ tomography characterisation}

	The experimental setup consists of a bare plate (Al2024 alloy with 0.6 mm thickness), a directed energy deposition system and an in situ synchrotron X-ray imaging experimental system, as shown in Fig.~\ref{fig_1}(a)(see `Methods' section for details). The chemical composition of the Al2024 alloy is given in Supplementary Table 1. The laser power(P), laser scanning speed (v), laser beam diameter (d) and laser scanning length with a single track used in this experiment were shown in Supplementary Table 2. During laser scanning, a stationary high-energy synchrotron X-ray beam (at the beamline 3W1 of Beijing Synchrotron Radiation Facility) penetrated through the specimen from horizontal direction. The transmitted X-ray beam carrying melt pool information was converted by a scintillator (LuAG:Ce) into visible light, which was recorded by a high-speed camera with a frame rate of 2000 Hz, and a spatial resolution of 4 \micro m per pixel. Therefore, all the physical dynamics were projected on a 2D imaging plane.

	\begin{figure*}[htbp]
		\center
		\includegraphics[width=1\linewidth]{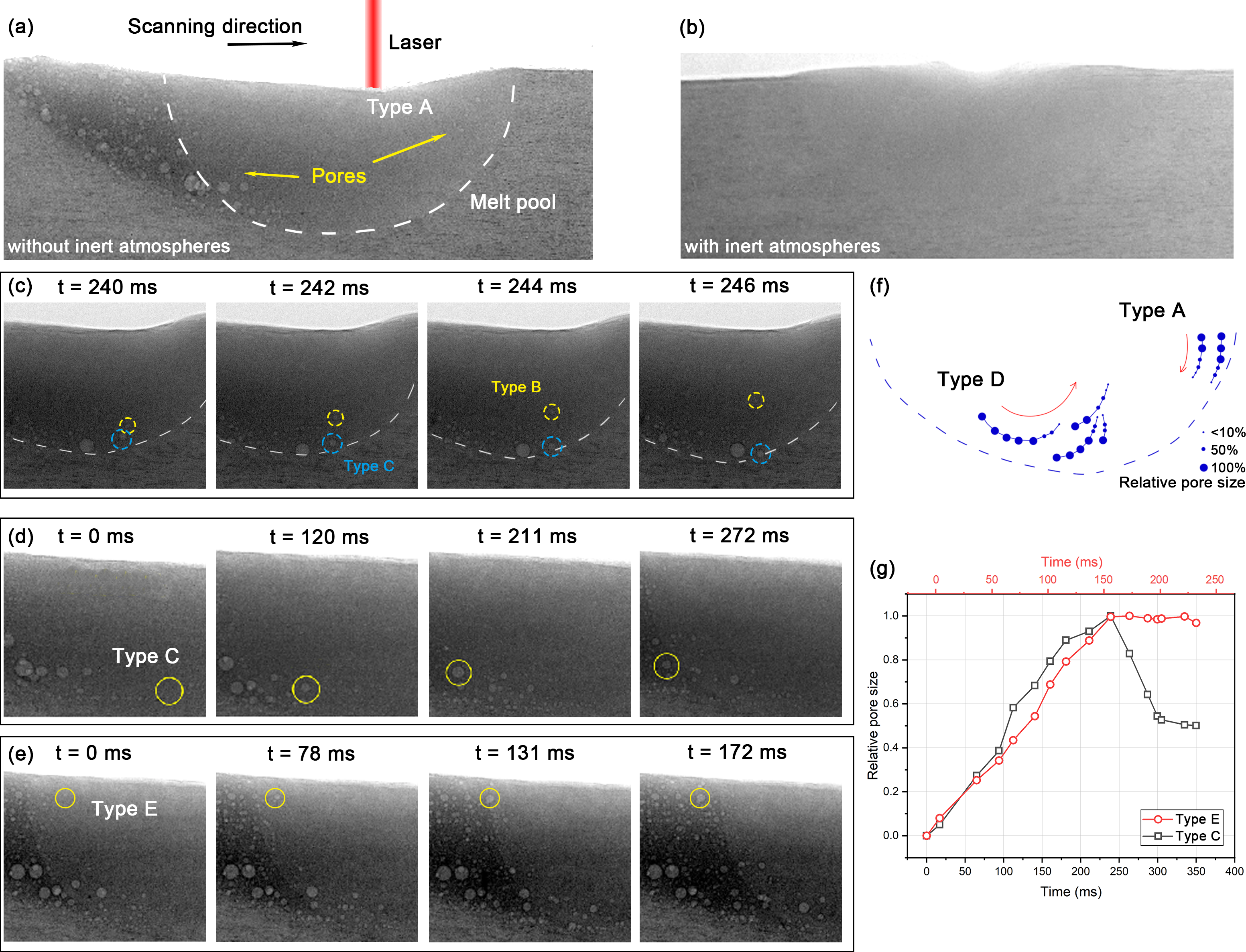}
		\caption{Comparison of remelting results with and without protective gas (a) without protective inert gas; (b) with protective inert gas. (c)-(e) Imaging data on the evolution of different types of pores over time in the melt pool. (f) Illustration of typical pore trajectories in the melt pool and relative volume evolution. (g) Evolution curves of average bubble volume during its life cycle for types C and E.}
		\label{fig_2}
	\end{figure*}	
	
	Subsequently, the formed sample are analyzed with X-ray computed tomography (CT) at Beamline 18U1 of the Shanghai Synchrotron Radiation Facility (SSRF) to quantify the characteristics of pores with 25 keV monochromatic x-rays. The voxel size after CT reconstruction is 1.6 $\mu$m. As seen in Fig.~\ref{fig_1}(c), $z$-axis of the $xyz$ coordinate system is along the build direction (BD) during additive manufacturing, and the $y$-axis is the direction of X-ray propagation. From the CT scans we can see that there are a large number of pores in the molded part, most of which are small pores (\textgreater 85\%) smaller than 4 \micro m, with only a few pores having a larger volume, up to several hundred microns. Since the spatial resolution in our {\it in-situ} X-ray imaging experiments was 4 \micro m, the 2D imaging data only shows the evolution of larger size pores. It can be seen that most of the pores are concentrated at the melt pool boundaries as well as at the upper surface, while relatively few are found in the inner regions of the melt pool.

	To characterise the initial and final states of the sample, we also used scanning electron microscope (SEM) and energy dispersive spectroscopy (EDS) to characterize the slices against the formed specimens. As shown in the bottom right of Fig.~\ref{fig_1} (c), combined with CT characterisation, it can be seen that the bubbles all maintain a more desirable spherical morphology. These {\it ex-situ} characterization results will be presented in the subsequent discussion section.

	\noindent \textbf{In situ synchrotron X-ray imaging during the DED process}

	Based on the observation of high-speed synchrotron X-ray imaging, it is obvious to see that bubbles are easy to be generated mainly in the environment without inert atmospheres, as shown in Fig.~\ref{fig_2} (a) and (b). The evolution of bubbles is mainly divided into three stages: bubble formation, growth and movement, and escape or capture by the solidification edge. According to the different behaviors of bubbles in the above stages, we classify bubbles into the following categories:

	\textbf{A. Formation at the front of the melt pool, movement towards the interior of the melt pool and gradual disappearance.} Fig.~\ref{fig_2} (a) moushows a two-dimensional image of the melt pool at one point in time without inert atmospheres. We can see that the bubbles generated at the melt pool front are smaller in size compared to bubbles in other areas. We use image recognition technology to track and analyze the trajectory and relative volume changes of these bubbles. Typical bubble evolution is shown in Fig.~\ref{fig_2} (f). We can see that the bubbles formed at the melt pool front move toward the center of the melt pool according to a certain arc trajectory, and in this process, the bubble volume decreases continuously and eventually disappears invisibly.

	\textbf{B. Formation at the bottom, movement towards the interior of the melt pool, gradual disappearance or escape from the surface.} The yellow circle in Fig.~\ref{fig_2} (c) marks a typical bubble, which forms at the bottom and rises rapidly along the center of the melt pool during $t$ = 240--246 ms. Its volume keeps getting smaller and eventually disappears. At the same time there are some bubbles that grow into larger bubbles at the bottom, which decrease in volume during the ascent, but maintain a certain volume when moving to the upper surface of the melt pool and escape, as can be seen in Fig.~S1 of the supplementary materials.

	\textbf{C. Formation at the bottom, movement towards the edge of the melt pool, and be captured by the solidification edge.} As shown by the bubble labeled by the blue circle in Fig.~\ref{fig_2} (c), it is formed at the bottom of the melt pool and has a short path of motion during $t$ = 240--246 ms, almost maintaining a stable position at the edge of the melt pool and not moving with the melt flow. Fig.~\ref{fig_2} (d) also shows a similar bubble's subsequent movement process, it can be seen that its position stabilizes the edge. With the melt pool movement, its relative position is constantly close to the back edge of the melt pool, and the volume gradually becomes larger. Finally it is captured by the solidification edge,  and then the volume gradually decreases and remains stable.

	\textbf{D. Formation at the back edge, movement towards the interior of the melt pool, gradual disappearance or escape from the surface}. There is a part of the bubble formed in the melt pool near the back edge, and moving with the melt flow, as shown in Fig.~\ref{fig_2} (f). These bubbles move along the arc trajectory to the center of the melt pool, while the volume continues to decrease. Similar to type B, bubbles are eventually disappeared or escaped from the surface.

	\textbf{E. Formation near the back edge, movement towards the edge, and be captured by the solidification edge.} As shown in Fig.~\ref{fig_2} (e), this type of bubble formed in the region near the back edge of the melt pool, and maintain a relatively stable position. Due to melt pool movement during the laser scanning process, bubbles gradually close to the back edge, and its volume gradually become larger. When it is captured by the solidification edge, the volume remains stable.

	\begin{figure}[htbp]
		\center
		\includegraphics[width=1\linewidth]{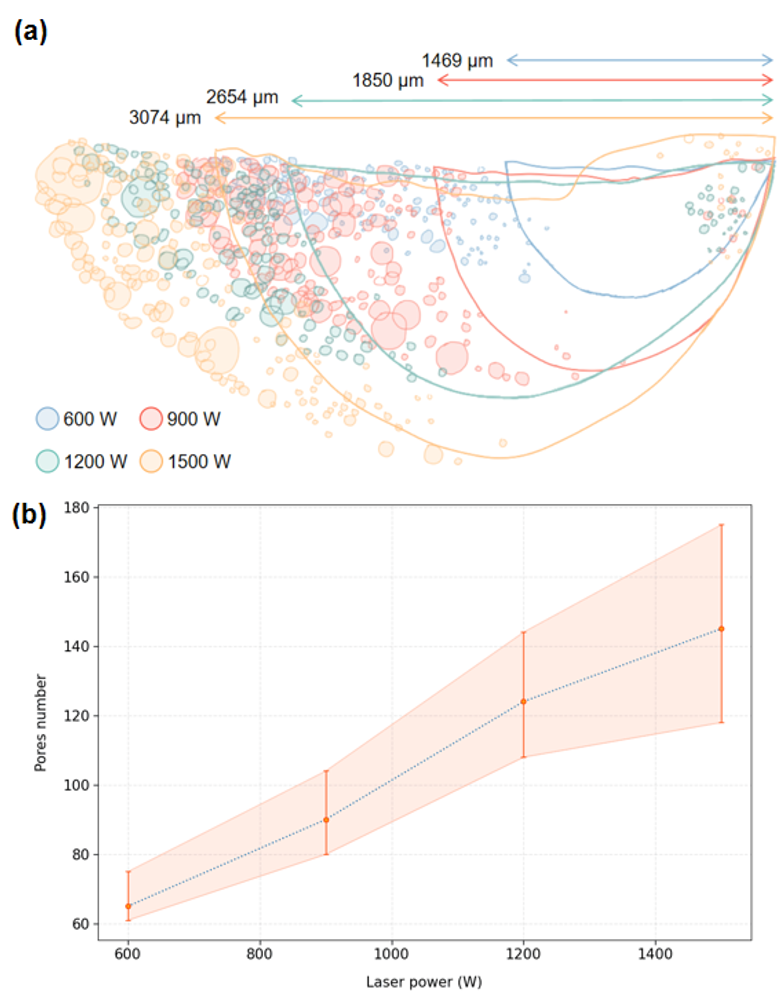}
		\caption{Dependence of melt pool and pores on laser powers. (a) Spatial distribution of pores for four different laser powers. (b) Number of pores as a function of laser power in the solidified area with the same volume.} \label{fig_3}
	\end{figure}

	\textbf{F. Bubbles entering from the surface of the melt pool.}	Due to the presence of external protective gases and metal powder impingement in the DED process, we observed that a small number of bubbles enter from the upper surface of the melt pool, move with the melt flow, and are either captured by the solidification edge or escape from the surface. These bubbles are shown in Fig.~S1 of the supplementary material.

	\textbf{G. Hollow area at the laser stop position.} At the end of the DED, as the laser stops, the melt pool area solidifies and shrinks rapidly, and we can observe that when the melt pool shrinks completely, a rapidly collapsing cavity region is formed at the top of the molded sample, as shown in Fig.~S2 of the Supplementary Material. Combined with the X-ray CT we can confirm that this is a hole that exhibits a crater shape.

	The seven types of bubbles described above contain common patterns of bubble evolution behavior in DED AM, with types C, E, and G ultimately remaining in the material and forming pore defects. While the other types do not lead to the formation of large pore defects. In order to investigate the effect of different laser parameters on the porosity, we conducted experiments with different laser powers, as shown in Fig.~\ref{fig_3}, which demonstrates the 2D shape extraction of instantaneous images at 600, 900, 1200, and 1500 W laser powers, respectively. It can be found that the higher the laser power, the larger the maximum bubble volume (as in Fig.~\ref{fig_3} (a)), and also the higher the number of pores per unit area (as in Fig.~\ref{fig_3} (b)). This indicates that bubbles of types C and E are significantly affected by the laser scanning strategy. It would be beneficial to reduce the porosity of DED-forming materials if the percentage of types C and E could be reduced so that more bubbles follow the evolutionary patterns of A, B, and D. Therefore, we will further delve into the physical mechanisms behind these different types of bubble patterns.

	\section{Discussion}
	\noindent \textbf{Pore deformation mechanism.}

	\begin{figure*}[hbp]
		\center
		\includegraphics[width=.7\linewidth]{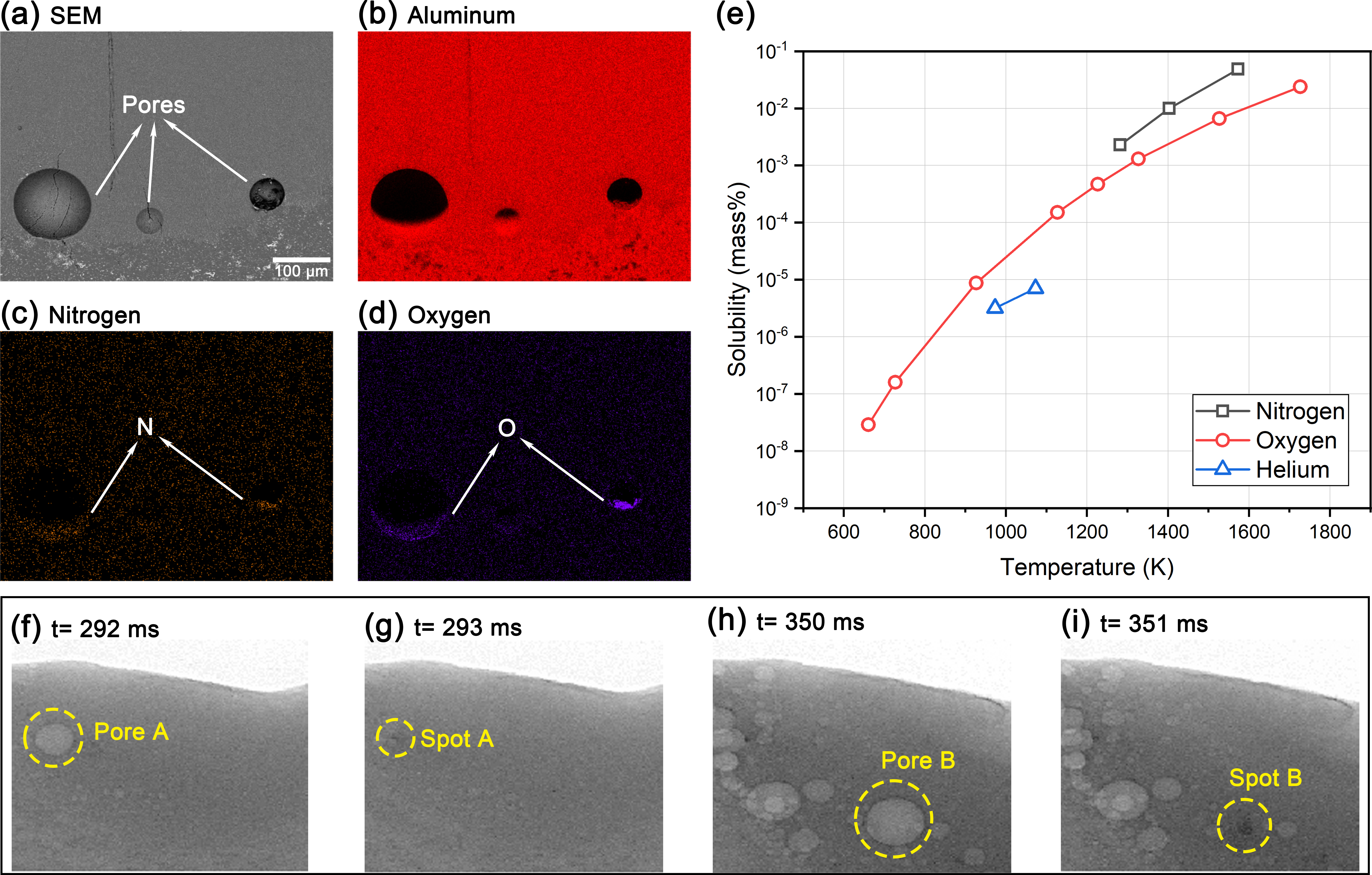}
		\caption{(a) SEM micrograph. (b--d) Corresponding EDS maps of elements Al, O and N as noted. (e) Elemental solubility in liquid aluminum as a function of temperature \cite{anyalebechi2022hydrogen,wriedt1986n}. (f-g) X-ray images showing collapse of pores A and B.}
		\label{fig_4}
	\end{figure*}

	From the experimental data, it is known that type C and type E are the main sources of a large number of pore defects during the scanning process. We tracked the volumetric evolution of these two types of bubbles separately over their full life cycle, as shown in Fig.~\ref{fig_3} (g). It can be seen that there is a significant difference in the volume evolution during the process of being captured by the solidification edge, where one stays constant while the other gets smaller. We hypothesize that this difference in the evolution pattern stems from the difference in the internal composition of their bubbles. For this reason we performed EDS analysis on the forming material to characterize the internal elemental composition of the bubbles. The results are shown in Fig.~\ref{fig_4} (a-d). It can be seen that among the three pores in (a), the edges of the two larger pores contain oxygen and nitrogen elements, while the small bubble in the middle does not contain these two types of elemental compositions. Oxygen and nitrogen are both major components of air, and at high temperatures oxygen will react chemically with aluminum to form oxides\cite{louvis2011selective}. Considering that these bubbles are generated during the DED process without protective inert gas, it is reasonable to assume that the ambient gas is closely related to the formation of the bubbles. The smaller pores that do not contain oxides and nitrides suggest a different class of bubble genesis. The edges of these bubbles contain only elements of aluminum alloy, so their formation and evolution should be entirely internal to the melt pool, with no contact with the external environment. Combining the observations of the kinematic patterns with the aforementioned imaging data, we give the following two main types of bubble formation and evolution mechanisms.

	\textbf{I. Dissolution and precipitation of ambient gases.} Due to the large volume of the DED process melt pool, its surface can be in full contact with the ambient gases, which are dissolved into the molten aluminum alloy by surface diffusion, or brought in by the impact of the powder, etc. The temperature inside the melt pool changes drastically due to the steep temperature gradient. The solubility of gases in liquid aluminum changes significantly with temperature (as shown in Fig.~\ref{fig_4} (e)). When the temperature is lowered from 1600 K to 800 K, nitrogen and oxygen solubility decrease by about five orders of magnitude\cite{anyalebechi2022hydrogen,wriedt1986n}. In solid aluminum alloys, the gas solubility is further reduced dramatically, almost to 0. When the dissolved gases move with the melt flow to the lower temperature region, they will precipitate to form bubbles, which will continue to gather and grow, and eventually be captured by the solidification edge.

	\textbf{II. Metal vapor bubbles.} Due to the high laser power, the aluminum alloy in the center of the melt pool is heated and vaporized (vaporization temperature is about 2740 K)\cite{desai1987thermodynamic}, and the resulting vapor recoil pressure is enough to overcome the surface tension of the liquid metal and the gravity, thus discharging part of the liquid metal, so that the molten pool at the laser action area is concave, forming a deep pit, i.e., the keyhole. Due to the external powder impact, the molten pool is disturbed and the keyhole is unstable. The bottom area is easy to collapse and detach, forming metal vapor bubbles, which are eventually captured by the solidification edge.

	According to the first type of mechanism, we can analyze the experimentally observed bubble collapse phenomenon. As shown in Fig.~\ref{fig_4} (f-i), there are some bubbles that collapse into a dense dot during the evolution process. X-ray imaging shows light black dots, which indicates that their x-ray transmittance is lower than that of the molten aluminum alloy. The transmittance of alumina is lower than that of liquid aluminum. Considering 25 keV X-ray, for example, the transmittance of alumina is 0.91, while that of liquid aluminum (900 K, density = 2.3 g/cm$^{3}$\cite{assael2006reference,smith1999measurement}) is 0.93 with the same thickness of 200 \micro m\cite{henke1993x}. This phenomenon confirms the existence of mechanism I.

	\begin{figure}[htbp]
		\center
		\includegraphics[width=.95\linewidth]{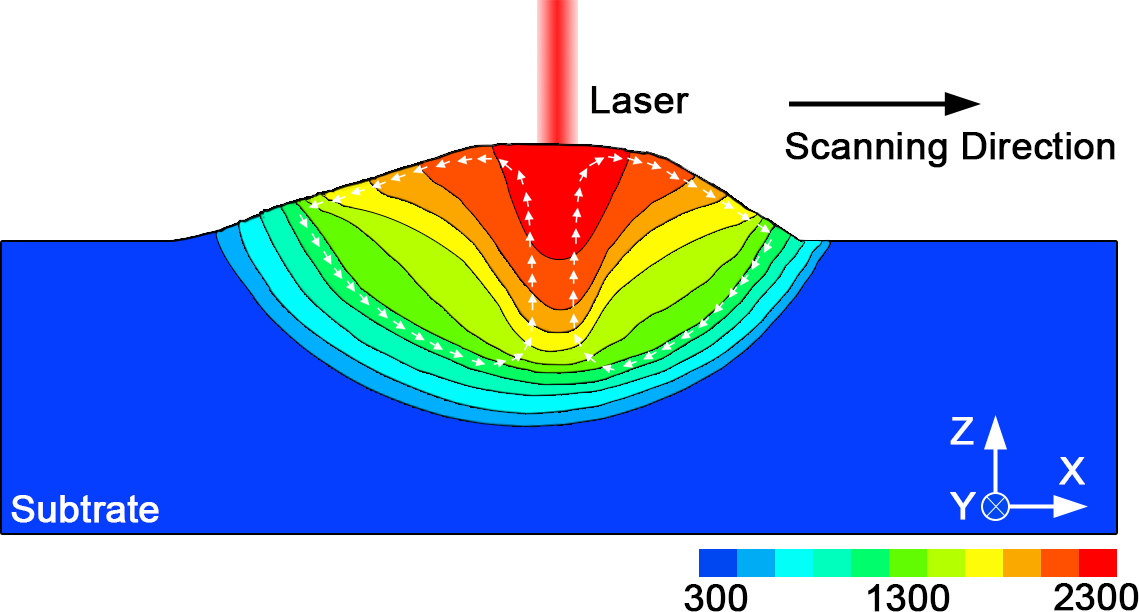}
		\caption{Two-dimensional temperature and flow fields obtained from the multiphysics coupled computational simulation.}
		\label{fig_5}
	\end{figure}

	In order to better analyze the motion process of this type of bubbles, we have carried out multi-physics coupled numerical simulations, and obtained the melt pool flow and temperature distribution results of the DED process by using the coupled flow and heat transfer model. The details of simulation are shown in the `Methods'. As shown in Fig.~\ref{fig_5}, the white arrows mark the main flow direction inside the melt pool, and some of bubbles will move along this direction, which is consistent with our experimental observation (Fig.~\ref{fig_2}). As the motion process proceeds, the temperature of the region where bubbles are located will change drastically, which will lead to different bubble evolution patterns. For the front region of the melt pool, when the gas moves clockwise with the melt flow, it first reaches the low temperature region of the front from the high temperature region above the center of the melt pool, and the gas precipitates to form bubbles. Subsequently, the gas moves again from the edge region to the center region. The temperature rises, and the gas dissolves again, while the bubble volume keeps getting smaller and disappears. This explains the bubble evolution pattern of type A. Also, Type D has a similar mechanism, where the dissolved gases continue to precipitate from the center of the melt pool into the low-temperature region, and then return to the center region again accompanied by dissolution and decreasing volume.

\begin{figure*}[htbp]
		\center
		\includegraphics[width=.8\linewidth]{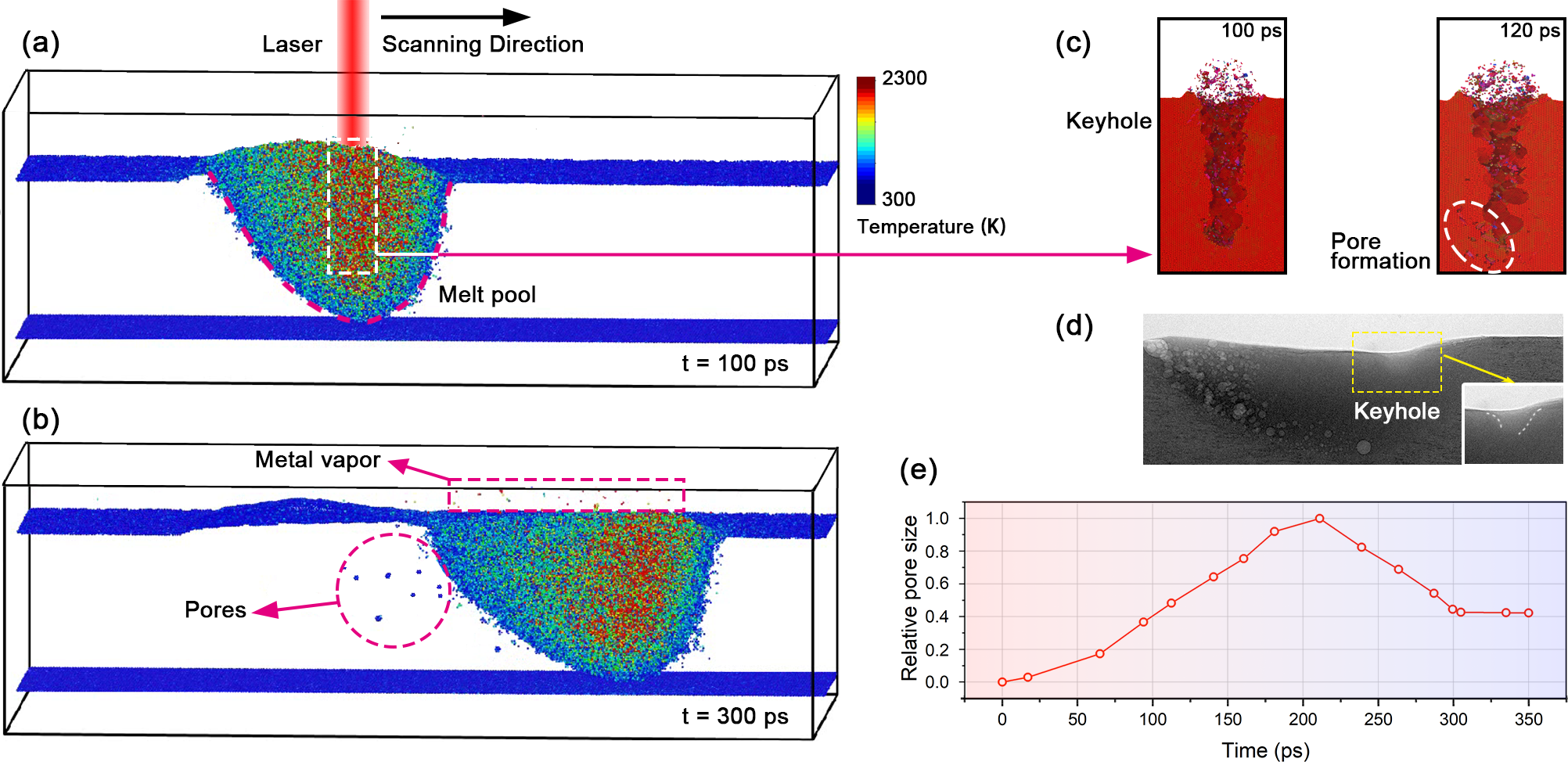}
		\caption{(a) and (b) presents the molecular dynamics simulation of the laser scanning process. As the scan progressed, many pores appeared behind the molten pool. (c) Partial magnification of the molten pool center. The figure shows the keyhole and the progress of the metal vapor bubble separation. (d) X-ray image of the top of the keyhole. (e) The curve of the average volume of the metal vapor bubble with time obtained by simulation.}
		\label{fig_7}
	\end{figure*}

	For the second type of mechanism, due to the presence of shock perturbations in the DED process and the limitation of the experimental contrast, it is difficult to directly observe a clear keyhole, and only few imaging data can be obtained in the top region of the keyhole, as shown in Fig.~\ref{fig_7} (d). This demonstrates the presence of keyhole. Since the keyhole is usually deep, this is why the bubble generation location for types B and C is below the melt pool. Where due to the influence received from the melt flow, the metal vapor bubbles will move upwards to the center of the melt pool and either reach the surface or may enter the keyhole again to escape. This explains the evolution of the type B. The other metal vapor bubbles are captured by the solidification edge and remain inside the material, as in type C. Also, the absence of any other element in the small pore in the EDS results corroborates the presence of metal vapor bubbles in the material.

\begin{figure*}[htbp]
		\center
		\includegraphics[width=1\linewidth]{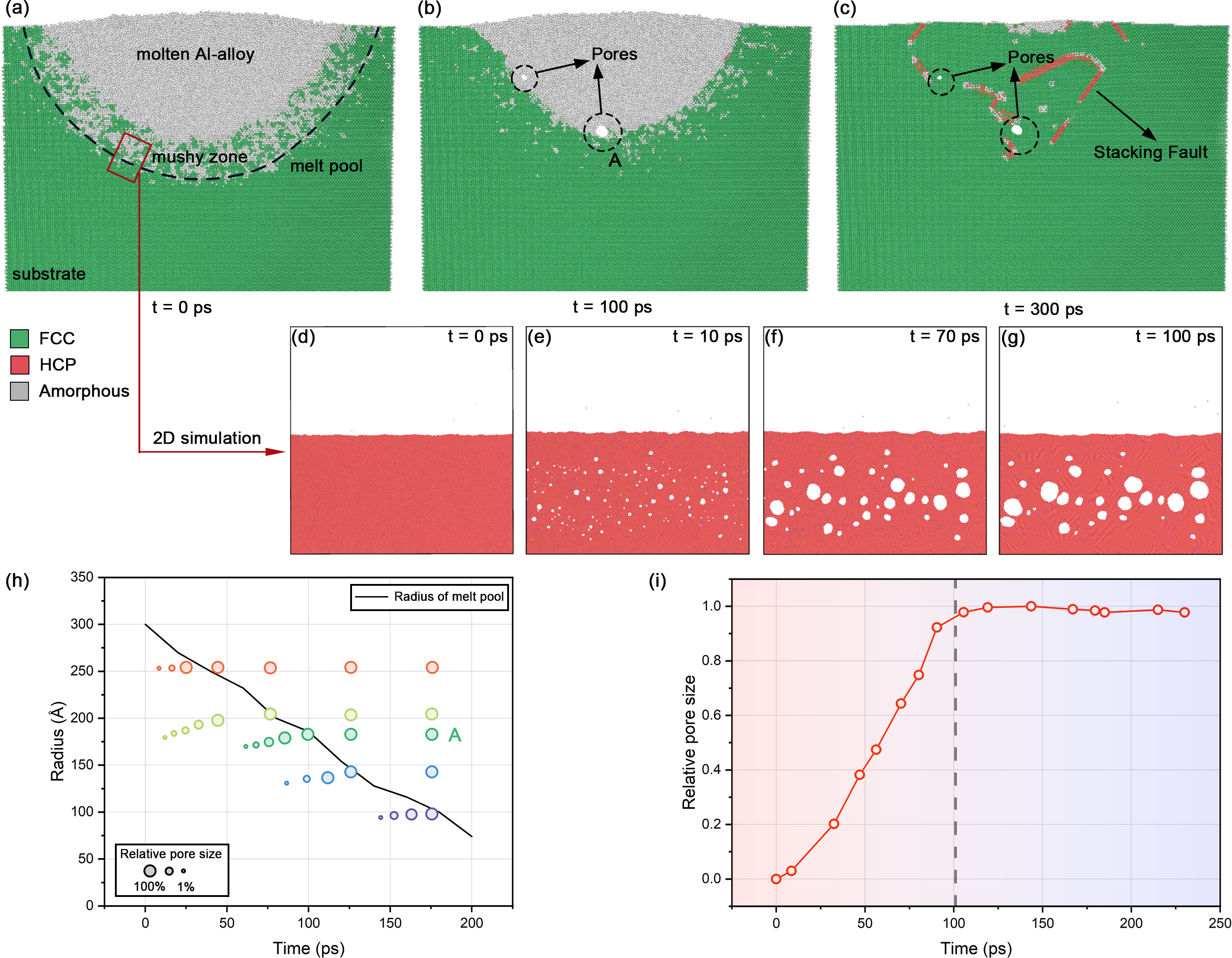}
		\caption{Simulation structures at different times. (a) The initial stage of solidification when the laser stops, t=0 ps; (b) The molten pool solidifies and shrinks, the interface advances upward, and some bubbles are produced, t = 60 ps; (c) The melt pool is completely solidified and the pores are fixed in the solid phase. (d)-(g) The structure of a two-dimensional molecular dynamics simulation of a local region at different times. (h) Evolution of pore volume and relative position with time in 3D simulation. (i) Evolution curve of pore volume with time in two-dimensional simulation.
		}
		\label{fig_6}
	\end{figure*}

	In order to further investigate the above mechanism, we carried out molecular dynamics simulations to investigate the roles of dissolved gases and metal vapors in the formation of pores from microscopic view. The simulation detail is described in the `Method' section. We approximate the temperature field obtained from the multiphysics simulation as a function and then apply it to the MD simulation to control the atomic temperature, so as to realize the melting and solidification process similar to the laser deposition process. Due to the limitation of the computational scale, we study a single-point melting model, as shown in Fig.~\ref{fig_6} (a-c), and set up the ambient gases, such as oxygen and nitrogen, above the melting pool. It can be seen that as the solidification proceeds, the melt pool continues to shrink and the temperature decreases, and bubbles have formed at the edge of the melt pool at $t$ = 100 ps, and by identifying the elemental compositions, it can be known that they are formed when the dissolved gases precipitate out of the molten metal. These bubbles are captured by the solidification edge and eventually remain inside the material to form porosity defects. We extracted and analyzed the bubble position and relative volume evolution of the simulation process, as shown in Fig.~\ref{fig_6} (h). It can be seen that the bubbles are basically formed close to the edge of the molten pool and become bigger, which is in line with the analysis of the aforementioned mechanism I. At the same time, their positions are almost constant, and they keep approaching the edge when the melt pool shrinks. We note that due to the drastic temperature gradient, the melt pool boundary is not sharp at the microscopic level, but there exists a solid-liquid mixing mushy zone, which hinders the motion of the bubbles with the heat flow to a certain extent due to the drag force. Theoretically larger bubbles are more likely to be bound in this region, while smaller bubbles are able to break away and enter the interior of the melt pool. Due to the finite computational size of simulation, it is difficult to have bubbles that are able to break away from this region, and therefore the bubbles formed in the simulation are all bound by the mushy region. This also explains why the bubbles of types C and E hardly move with the melt flow and are adsorbed at the edge of the melt pool.

	In order to further statistically analyze the evolution of dissolved bubbles, we constructed a two-dimensional model to carry out molecular dynamics simulations for a localization, as shown in Fig.~\ref{fig_6} (d-g), with the ambient gas on the top and the melt on the bottom, and approximated temperature evolution at the top of the molten pool as a function over time from the multiphysics simulation and apply it to the solidification process. By counting the volume of bubbles precipitated, as shown in Fig.~\ref{fig_6} (i), we can see that their volume keeps getting larger with decreasing temperature, and eventually remains in a relatively stable state after solidification. This is basically consistent with the volume evolution of type E. The reliability of mechanism I is further verified.

	In addition to this, we simulated the laser scanning process using MD. By setting the appropriate scanning speed and laser power, the laser scanning process can be simulated on a microscopic scale. As shown in Fig.~\ref{fig_7}, it can be seen that when the laser power is high, metal vapor is generated inside, forming an irregular keyhole. As the scanning proceeds, the disturbance inside the melt pool is intense, and it is difficult for the keyhole to remain stable. Thus some metal vapor bubbles detach, as shown in Fig.~\ref{fig_7} (c). These bubbles are captured by the solidification edge and remain inside the solidified body, forming a porosity defect. We traced the evolution of this bubble, and its volume evolution and motion trajectory are similar to that of type C, which aids in verifying the reliability of mechanism II.

	In addition, for the holes of type G, according to the simulation, it can be seen that the molten pool keeps shrinking during the condensation process due to the laser stopping, while the surface area cools down faster due to the direct contact with the external environment, as shown in Fig.~S4 of the supplementary material. At the same time, the aluminum alloy is in contact with air, which is easy to form a layer of aluminum oxide film on the surface. These two  result in the presence of a small amount of unsolidified liquid aluminum at the top of the melt pool at the end of the solidification period, wrapped around the solidified surface. With further cooling, liquid and solidification shrinkage proceed, whereupon the melt pool collapses to form a hole with a certain degree of vacuum. If the solid thin layer at the top cannot withstand the stress of atmosphere, it ruptures, as observed in this experiment. This type of holes can be avoided by reciprocal remelting scans since they form only in the laser stop region.

	By analyzing the above pores formation mechanism and the evolution mode of different types of bubbles, combined with the experimental data of different laser parameters, we can propose some methods to reduce the porosity. First of all, the use of low solubility inert gases as a protective gas can significantly reduce the formation of bubbles, as can be seen from Fig.~\ref{fig_4} (e), helium solubility in the aluminum melt is much smaller than oxygen and nitrogen, and argon is about 3 orders of magnitude less soluble than helium. So low solubility can be considered as basically no gas dissolution. At the same time inert gases are chemically stable, even at high temperatures, it is difficult to react with the aluminum melt. Secondly, the sharp temperature gradient leads to a wider solid-liquid mixing zone, which affects the movement of the bubbles with the melt flow and is not conducive to the escape of the bubbles. Finally, a larger melt pool allows a fuller environment for bubbles to evolve and grow, resulting in larger bubbles that are more difficult to break away from the edge of the melt pool and produce larger pores. Therefore a lower laser power with a stable scanning speed may be more appropriate for reducing porosity in DED.

	In summary, we performed {\it in situ} real-time synchrotron X-ray imaging diagnostics of the laser energy deposition process of aluminum alloys such as Al2024 at the 3W1 beamline of the BSRF, and combined with multiphysics coupled computation and molecular dynamics simulations to analyze the evolution process of seven different types of bubbles. We found that solubility differences and metal vapors are the two main mechanisms for bubble formation inside the melt pool, which are affected by the laser process, both of which play a dominant role in different types of bubbles, respectively. At the same time, due to the intense temperature gradient, the formation of solid-liquid mixing zone and the melt flow play a joint role in determining the evolution of bubble movement in the melt pool. These mechanisms are closely related to the laser power. Therefore, according to our results, the selection of inert gases such as argon as the ambient gas, appropriately reduce the laser power and maintain a gentle scanning speed can inhibit the formation of pores in DED process and promote the escape of bubbles, which can effectively improve the molding quality.

	\section*{Data availability}
	
	Raw data were generated at the Beijing Synchrotron Radiation Facility. The main data needed to supporting our findings are provided in the manuscript or the supplementary materials. Other reference data and code can be provided on request.

	\bibliographystyle{elsarticle-num}

	\bibliography{main.bib}

	\clearpage
	
\end{document}